\documentclass[aps, prb, preprint, amsmath,amssymb,showpacs,superscriptaddress]{revtex4-1}

\usepackage{graphicx}
\usepackage{color}
\usepackage{amsthm}
\usepackage{amsmath}
\usepackage{amssymb}
\usepackage{setspace}
\usepackage{braket}
\usepackage{float}
\renewcommand{\(}{\left(}
\renewcommand{\)}{\right)}

\makeatletter
\@ifundefined{textcolor}{}
{
 \definecolor{BLACK}{gray}{0}
 \definecolor{WHITE}{gray}{1}
 \definecolor{RED}{rgb}{1,0,0}
 \definecolor{GREEN}{rgb}{0,1,0}
 \definecolor{BLUE}{rgb}{0,0,1}
 \definecolor{CYAN}{cmyk}{1,0,0,0}
 \definecolor{MAGENTA}{cmyk}{0,1,0,0}
 \definecolor{YELLOW}{cmyk}{0,0,1,0}
}

\renewcommand{\H}{\mathcal{H}}

\begin{document}

\title{Spectroscopic evidence for type II Weyl semimetal state in MoTe$_2$}

\author{Lunan Huang}
\affiliation{Ames Laboratory, U.S. DOE and Department of Physics and Astronomy, Iowa State University, Ames, Iowa 50011, USA}

\author{Timothy M. McCormick}
\affiliation{Department of Physics and Center for Emergent Materials, The Ohio State University, Columbus, OH 43210, USA}

\author{Masayuki Ochi}
\affiliation{Department of Physics, Osaka University, Toyonaka, Osaka 560-0043, Japan}

\author{Zhiying Zhao}
\affiliation{Department of Physics and Astronomy, University of Tennessee, Knoxville, Tennessee 37996, USA}

\author{Michi-to Suzuki}
\affiliation{RIKEN Center for Emergent Matter Science (CEMS), Wako, Saitama 351-0198, Japan}

\author{Ryotaro Arita}
\affiliation{RIKEN Center for Emergent Matter Science (CEMS), Wako, Saitama 351-0198, Japan}
\affiliation{JST ERATO Isobe Degenerate $\pi$-Integration Project, Advanced Institute for Materials Research (AIMR), Tohoku University, Sendai, Miyagi 980-8577, Japan}

\author{Yun Wu}
\affiliation{Ames Laboratory, U.S. DOE and Department of Physics and Astronomy, Iowa State University, Ames, Iowa 50011, USA}

\author{Daixiang Mou}
\affiliation{Ames Laboratory, U.S. DOE and Department of Physics and Astronomy, Iowa State University, Ames, Iowa 50011, USA}

\author{Huibo Cao}
\affiliation{Quantum Condensed Matter Division, Oak Ridge National Laboratory, Oak Ridge, Tennessee 37831, USA}

\author{Jiaqiang Yan}
\affiliation{Materials Science and Technology Division, Oak Ridge National Laboratory, Oak Ridge, Tennessee 37831, USA}
\affiliation{Department of Materials Science and Engineering, University of Tennessee, Knoxville, Tennessee 37996, USA}

\author{Nandini Trivedi}
\email[]{trivedi.15@osu.edu}
\affiliation{Department of Physics and Center for Emergent Materials, The Ohio State University, Columbus, OH 43210, USA}

\author{Adam Kaminski}
\email[]{kaminski@ameslab.gov}
\affiliation{Ames Laboratory, U.S. DOE and Department of Physics and Astronomy, Iowa State University, Ames, Iowa 50011, USA}

\date{\today}

\begin{abstract}
\bf In a type I Dirac or Weyl semimetal, the low energy states are squeezed to a single point in momentum space when the chemical potential $\mu$ is tuned precisely to the Dirac/Weyl point. Recently, a type II Weyl semimetal was predicted to exist, where the Weyl states connect hole and electron bands, separated by an indirect gap. This leads to unusual energy states, where hole and electron pockets touch at the Weyl point. Here we present the discovery of a type II topological Weyl semimetal (TWS) state in pure MoTe$_2$, where two sets of WPs $(W_2^\pm, W_3^\pm)$ exist at the touching points of electron and hole pockets and are located at different binding energies above E$_F$. Using ARPES, modeling, DFT and calculations of Berry curvature, we identify the Weyl points and demonstrate that they are connected by different sets of Fermi arcs for each of the two surface terminations. We also find new surface ``track states" that form closed loops and are unique to type II Weyl semimetals. This material provides an exciting, new platform to study the properties of Weyl fermions.

 \end{abstract}
\maketitle

\section{INTRODUCTION}

It is quite surprising and yet exhilarating that non-interacting or quadratic Hamiltonians can continue to provide so much richness from graphene, to topological insulators and topological superconductors. This list was recently expanded by discovery of topological Weyl semimetals (TWS), the relatively robust three-dimensional analogs of graphene. With all three Pauli matrices involved in the Hamiltonian, perturbations only shift the position of the node in momentum space but do not open a gap. 

While the massless solution to the Dirac equation~\cite{Weyl1929} was first proposed by Hermann Weyl in 1929, there are no known examples of Weyl fermions in particle physics. Quantum materials' analogs have been proposed in various classes of topological Dirac and Weyl semi-metals \cite{Hsieh2008,PhysRevLett.107.127205,Liu864,Neupane2014} where a pair of Dirac nodes can be separated into two Weyl points (WPs) by breaking either inversion or time reversal invariance. The topological nature of a TWS is reflected in the Berry fluxes of opposite chirality circulating around the WPs and the presence of a Fermi arc formed between the projections of the two Weyl points on a surface at which the bulk is truncated.

Recently, two types of TWS have been identified: Type I TWS can be understood as the limiting point of a semiconductor with a direct band gap that closes linearly at a set of isolated points. 
As a consequence, there is zero density of states if the chemical potential is tuned to the energy of the WPs. Type I TWS have been observed in the TaAs family (TaAs, NbAs and TaP)~\cite{Xu07082015,Xue1501092,Xu2015a}, and also predicted to occur in pyrochlore iridates~\cite{Huang2015, doi:10.7566/JPSJ.84.073703, Wan2011}. Type II TWS, on the other hand, can be understood as the limiting point of an indirect gap semiconductor that evolves into a compensated semi-metal with electron and hole pockets that touch at a set of isolated points with a finite density of states at the chemical potential. The two WPs connected by a Fermi arc need not occur at the same energy. MoTe$_2$, WTe$_2$ and SrSi$_2$ are predicted to be such a type II TWS~\cite{Sun2015, Soluyanov1507,2015arXiv150305868H}. There are some signatures of such a state in mixed compound Mo${_{0.45}}$W${_{0.55}}$Te$_2$\cite{Belopolski2015}. Here we present the first evidence for type II Weyl semimetal behavior in the stochiometric, low scattering material such as MoTe$_2$.

One of the most exciting properties of a TWS is the existence of gapless Fermi arcs on the surface. A Fermi surface, defined as the locus of gapless excitations, is typically a closed contour that separates filled states from empty states at zero temperature. In view of that, a chopped up Fermi surface with the two pieces on opposite surfaces is a novel state of matter. Surface sensitive probes such as angle-resolved photoemission spectroscopy have a decided advantage in investigating the structure of arcs, connectivity of electron and hole pockets and locations of Weyl points, which is the topic of our paper. 

\section{Model for type II TWS}

\begin{figure*}
	\centering
	\includegraphics[width=\columnwidth]{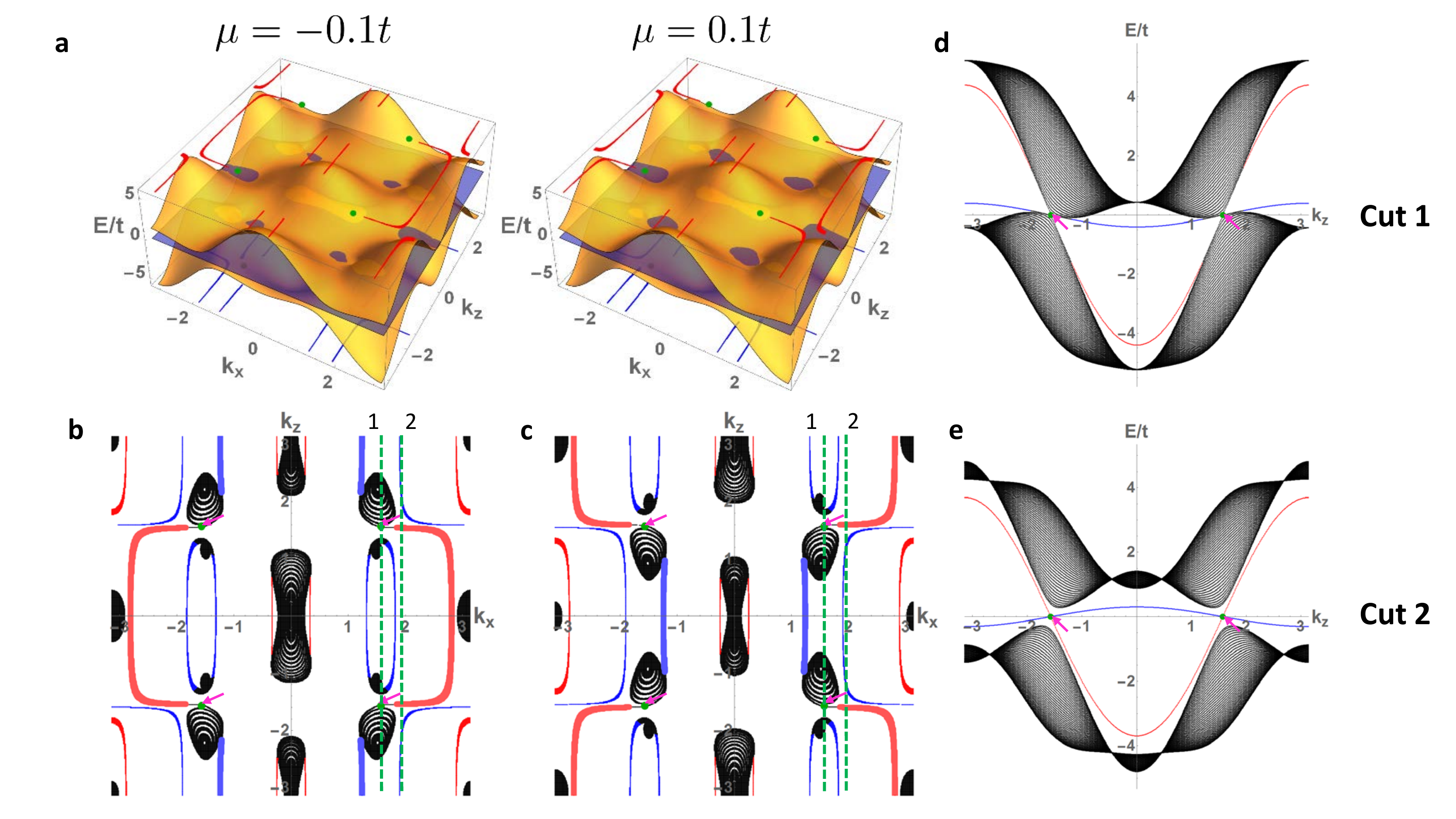}
	\caption{Simple model of type II Weyl semimetal described by a two band model given by Eq. \ref{hinvkernel} which exhibits four Weyl nodes. 
	{\bf a} Electronic band structure for $\mu = \pm 0.1t$ indicated by the blue translucent plane. 
	{\bf b,c} The topological surface states and Fermi arcs on surface A (in red) and B (in blue) are calculated for a slab geometry confined along the $y$-direction. The bulk bands are shown in black. When $\mu$ = 0 exactly, the electron and hole pockets touch and the arcs terminate on the node (green dot) itself. For Fermi energy below (above) the nodal energy, arcs of surface states connect the Fermi hole (electron) pockets surrounding a node rather than terminating on a node. 
	{\bf d,e} Energy dispersion along $k_z$ at fixed $k_x$ as shown by cuts in panels (b, c). Cut 1 along $k_x$ = $\pi/2$ shows the bulk electron and hole bands touching at the node and the merging of surface states into the bulk {\em away} from the Weyl node. Cut 2 along $k_x$ = 0.63$\pi$ shows a gap between the bulk bands and a surface state that disperses with opposite velocities at the projections of the two WPs. The WPs are located at $(k_x,k_z)$ = $(\pm \pi/2,\pm\pi/2)$ indicated by pink arrows pointing to green dots.}
	\label{fig1}
\end{figure*}

To set the stage for interpretation of the experimental results, we investigate a two-band lattice model which breaks inversion symmetry but is invariant under time-reversal symmetry. The main lessons learned by examining this model are shown in Fig. \ref{fig1} and summarized here:
(1) The minimum number of four Weyl nodes in this type II TWS occur at $E=0$ at the touching point of electron and hole pockets
in contrast with a type I TWS that has a zero density of states at $E=0$. The touching of electron and hole bands in our model is similar to the touching of the electron and hole bands in the experimental data shown in Fig. \ref{fig2}a and b.
(2) For a slab geometry, constant energy cuts at $E=0$ show Fermi arcs on surface termination A and B that connect Weyl points of opposite chirality.
In addition there are what we term ``track states" that exist on the surface and pass through the WPs but, unlike Fermi arcs, form closed loops. For $E<0$, the projections of the WPs are within the hole pocket, and at the surface the arc states connect the two hole pockets and the track states loop around the electron pockets. The opposite is true for $E>0$.
(3) The energy dispersion clearly shows a surface state dispersing separately from the bulk bands and merging with the bulk bands close to the WP in Fig. \ref{fig1}d. This is corroborated by the experimental data around the Weyl nodes in Fig. \ref{fig2}i and Fig. \ref{fig3}n where the arc merges with the bulk states.

We consider the following Hamiltonian for a two-band lattice model which breaks inversion symmetry and is invariant under time-reversal:

\begin{equation}
\hat{H}_{\textrm{Inv}} = \sum_{\mathbf{k}} \hat{c}^{\dagger}_{\mathbf{k}\alpha} \( \hat{\H}(\mathbf{k}) \)_{\alpha \beta} \hat{c}_{\mathbf{k}\beta}
\label{invHam}
\end{equation}

where $\hat{c}^{(\dagger)}_{\mathbf{k}\alpha}$ annihilates (creates) an electron at momentum $\mathbf{k}$ in orbital $\alpha$ and

\begin{multline}
\hat{\H}(\mathbf{k}) =\gamma (\textrm{cos}(2 k_x)-\textrm{cos}(k_0))(\textrm{cos}(k_z)-\textrm{cos}(k_0))\hat{\sigma}_{0}\\- (m(1-\textrm{cos}^{2}(k_z)-\textrm{cos}(k_y))+2t_x(\textrm{cos}(k_x)-\textrm{cos}(k_0)))\hat{\sigma}_1
\\-2t\ \textrm{sin}(k_y) \hat{\sigma}_2-2t\ \textrm{cos}(k_z) \hat{\sigma}_3 .
\label{hinvkernel}
\end{multline}

Here $\hat{\sigma}_{i}$ is the $i$-th Pauli matrix for $i = 1,2,3$ and $\hat{\sigma}_{0}$ is the $2\times2$ identity matrix. This model has four Weyl nodes located at $E = 0$ and $\mathbf{k} = (\pm k_0, 0, \pm \pi/2)$. The term in $\hat{\H}(\mathbf{k})$ proportional to $\hat{\sigma}_{0}$ produces a uniform shift in both energy bands. Such a momentum-dependent shift will result in a non-vanishing density of states from electron and hole pockets which touch at the Weyl node and a tilt of the Weyl nodes characteristic of a type II TWS. Henceforth, we set the parameters $m = 2t$, $t_x = t/2$, $k_0 = \pi/2$, and $\gamma = 2.4t$. The bulk band structure for this parameter choice can be seen in Fig. \ref{fig1}a which shows hole and electron pockets touching at the Weyl nodes as well as pockets disconnected from the nodes. Similar Fermiology is also present in the MoTe$_2$ system and we can gain insight into this and other related materials by taking advantage of the lattice model's simplicity and tunability.

We examine the structure of the surface state configuration by considering the model in Eq. (\ref{invHam}) in a slab geometry finite in the $y$- direction with $L$ layers but infinite in the $x$- and $z$-directions. We label the states as ``surface termination B" (``surface termination A") if they are exponentially localized at $\braket{y} = 1$ ($\braket{y} = L$). Fig. \ref{fig1} also shows the surface states at $\mu = \pm 0.1t$ overlaid on the bulk band structure. We show constant energy cuts through the band structure of the slab geometry in Fig. \ref{fig1}b and c for $\mu = \pm0.1t$. When $\mu < 0$, the projections of the Weyl nodes (shown by green dots) are enclosed by hole pockets. Each of these hole pockets are connected to another pocket containing a node of opposite chirality by one Fermi arc on surface A (B) shown as a thick light red (blue) line. When $\mu > 0$, the projections of the Weyl nodes are enclosed by electron pockets which are similarly connected by Fermi arcs on the surfaces. At precisely $\mu = 0$, because all of the nodes lie at $E = 0$, all Fermi arcs terminate on the nodes themselves as in a type I TWS.

The slab configuration energy dispersion for fixed $k_x$ is shown in Fig. \ref{fig1}d and 1e. These cuts are shown as green dashed lines labeled cut 1 and cut 2 respectively. We can see that at the Weyl nodes, the red surface bands in 1d disappear into the bulk. As we move past the Weyl points in e, we see that these two red bands combine into a single continuous band.

\section{ARPES Results}

MoTe$_2$ is a semimetal that crystallizes in a orthorhombic lattice. The Fermi surface of MoTe$_2$ also has two 2-fold symmetry axes, along $\Gamma$ - X and $\Gamma$ - Y directions. The lattice constants are $a$ = 6.33 \AA, $b$ = 3.469 \AA. Due to breaking of the inversion symmetry there are two different possible terminations of the cleaved sample surface, referred to as termination ``A" and ``B" respectively. The two different terminations also have different surface band structures as seen by laser-based angle resolved photoemission spectroscopy (ARPES) and corroborated by DFT calculations.

We identify electron and hole bands in the spectroscopic data shown in Figs. \ref{fig2} and \ref{fig3}. The hole bands at the center of the Brillouin zone have a ``butterfly" shape. The electron pockets shaped like ovals are located on each side of the butterfly. There are also two banana like hole pockets partially overlapping the oval electron pockets. The configuration of these pockets can be seen at the Fermi energy in Fig. \ref{fig2}a and 10 meV above the Fermi energy in Fig. \ref{fig2}b and their electron or hole character is easily identified because hole (electron) pockets shrink (expand) with increasing energy. A simplified sketch of constant energy contours of electron and hole bands is shown in Fig. \ref{fig2}c.

The central hole pocket touches the electron pockets at four Weyl points shown as red dots in Fig. \ref{fig2}a-c which we label as $W_2$. The outer banana shaped hole pockets also touch the oval electron pockets at two other Weyl points labeled as $W_3$. At surface termination A, Fig. \ref{fig2}b, those two types of Weyl points are connected by topological arcs seen as white-gray high intensity areas. For this surface termination there is no strong evidence for arcs connecting positive and negative chirality $W_2$ nor positive and negative chirality $W_3$ points. The situation for surface termination B is more complicated as shown in Fig. \ref{fig2}d. There seems to be a sharp contour connecting both sets of $W_2$ and $W_3$ points. Most likely this is a track state discussed above. The examination of constant energy plot at energy of 30 meV below $E_F$ (Fig \ref{fig2}e), reveals that there are actually two bands present. In addition to the track state, there is also an arc present that connects positive and negative chirality $W_2$ points. Although present data does not allow us to definitely demonstrate a connection between positive and negative chirality $W_3$ points, we can deduce that they are likely connected, so the arcs on surface A between $W_2$-$W_3$ together with arcs on surface B $W_2^+$-$W_2^-$ and $W_3^+$-$W_3^-$ form a closed loop when connected via the bulk of the sample.

We now examine the locations of the Weyl points in the band dispersion. In Fig. \ref{fig2}f-i we plot the band dispersion along $k_y$ cut for selected values of $k_x$. At $k_y$=0.36 $\pi/b$ (panel f) two bands are clearly visible: an ``M" shaped band at higher binding energy and a ``U" shaped band at slightly lower binding energy. Both bands appear connected at zero momentum with Dirac-like structure. As we move towards the zone center, both bands move to lower binding energy and their energy separation decreases. In panel h, the tips of the ``M" shaped band (red dotted line) touches the $E_F$ and form parts of the butterfly hole pockets. As these tips move above $E_F$, they touch merge with wings of the ``U" shaped electron band (white dotted line) forming two Weyl points approximately 20 meV above $E_F$ marked by black dots. At each side of the symmetry line, they form two tilted cones characteristic of a type II Weyl node. The data along $k_x$ direction are shown in Fig. \ref{fig3}d-o along with results of calculations (Fig. \ref{fig3}p-w) for the two surface terminations. The surface termination A is characterized by lower binding energy of electron pocket in panels d-g, when compared to the data from surface termination B shown in panels h-k and l-o. The data in panels l-o best illustrates the formation of the $W_2$ points. In panel l, the hole band is marked with red dashed line, while the electron band is marked with white dashed line. As we move away from the symmetry line, the separation between those bands becomes smaller and they merge at a point located $\sim$20 meV above $E_F$ marked by red dot in panel n. For higher values of $k_y$ momentum they separate again as seen in panel o. The DFT calculation also demonstrates the energy difference of the band locations for the two terminations and formation of the $W_2$ Weyl point that agrees with experiment on a qualitative level.

The momentum location of the experimentally determined Weyl points is somewhat different from DFT predictions (marked as pink dots in \ref{fig2}a,b) most likely due to high sensitivity of the band calculation to structural parameters. Table I summarizes the positions of WPs determined from experiment and DFT. Despite the discrepancy between the predicted locations of the Weyl nodes from DFT and where they are located experimentally, in each case they are at the touching points of the electron and hole bands. In the $k_y$ = 0 cuts shown in Fig. \ref{fig3}d,h,l,p,t, band 1 is connected to bulk states below the Fermi level, while band 3 dips down and goes into bulk just before it reaches the Weyl point. As we increase $k_y$, band 1 and band 3 merge together. In the $k_y$ = 0.1$\pi/a$ cuts, the two bands merge into one band which goes through the position of the projection of $W_2$. This behavior is exactly the the behavior predicted in Fig. \ref{fig1}d and \ref{fig1}e.

\section{DFT and Topological Analysis}
\begin{figure*}
	\centering
	\includegraphics[width=\columnwidth]{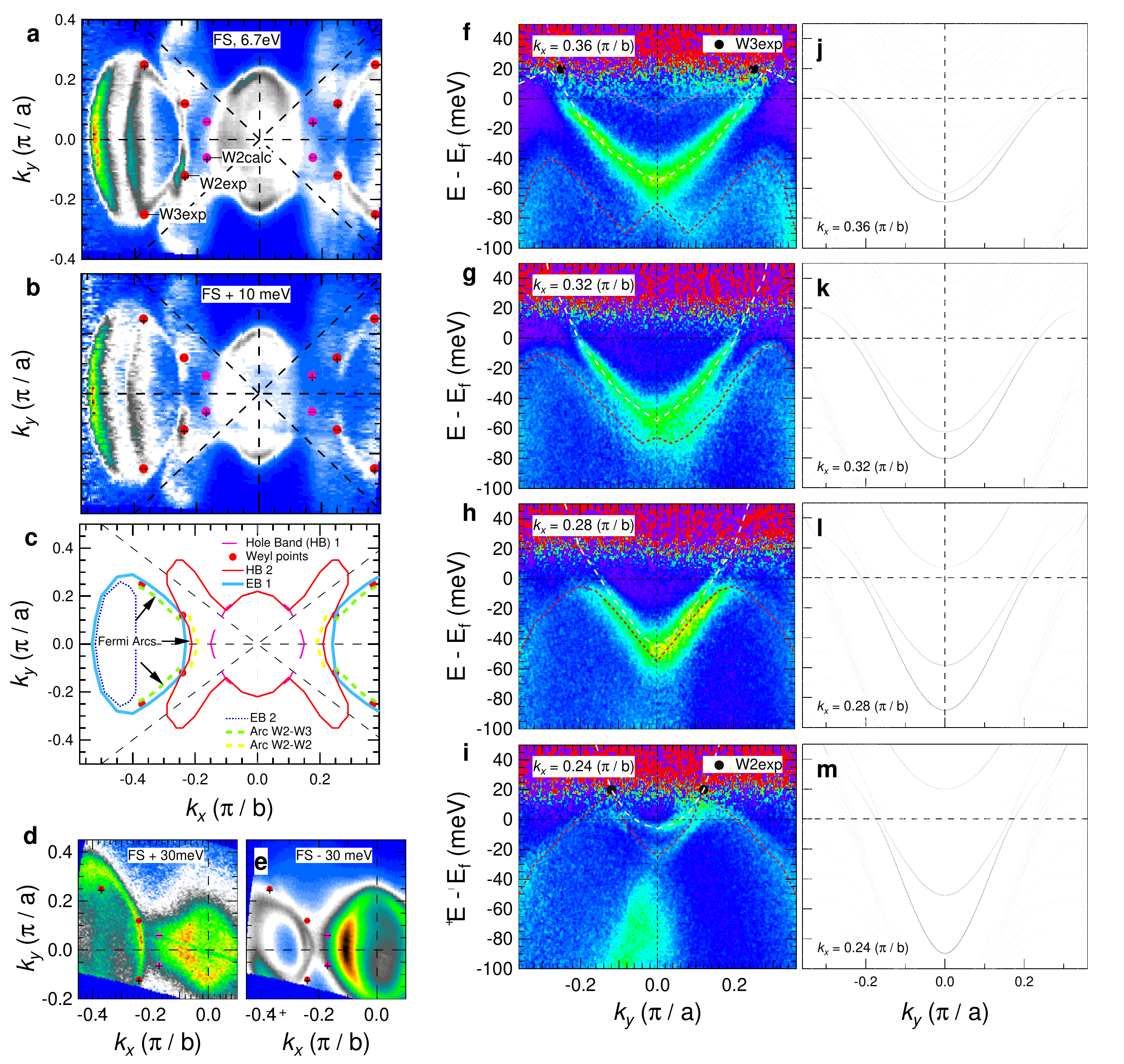}
	\caption{Experimental Fermi surface and band structure of MoTe$_2$. 
	{\bf a} Constant energy intensity plot measured at $E_F$ using 6.7 eV photons for a sample with termination A. The calculated (DFT) positions of Weyl points $W_2$ are marked as pink dots, while experimentally determined locations of $W_2$ and $W_3$ points are marked as red dots. The chiralities of Weyl points are marked with ``+" and ``-" and their locations $(k_x,k_y,E)$ are summarized in Table \ref{weylLocationTable}.
	{\bf b} Same as in {\bf a} above but taken at 10 meV above $E_F$.
	{\bf c} A sketch of constant energy contours of electron and hole bands showing the locations of Weyl points and Fermi arcs. 
	{\bf d} Constant energy contour measured at 30 meV above $E_F$ using 5.9 eV photons for a sample with termination B. Positions of calculated and measured Weyl points are marked as above.
	{\bf e} Same surface termination and photon energy as {\bf d} but at 30 meV below $E_F$. 
	{\bf f - i} Experimental band dispersion along cuts at $k_x$ = 0.24, 0.28, 0.32 and 0.36 $\pi/b$.
	{\bf j - m} Calculated band dispersion for a sample with termination A along $k_x$ = 0.24, 0.28, 0.32 and 0.36 $\pi/b$. Bands plotted with darker lines have more surface weights.}
	\label{fig2}
\end{figure*}

\begin{figure*}
	\centering
	\includegraphics[width=\columnwidth]{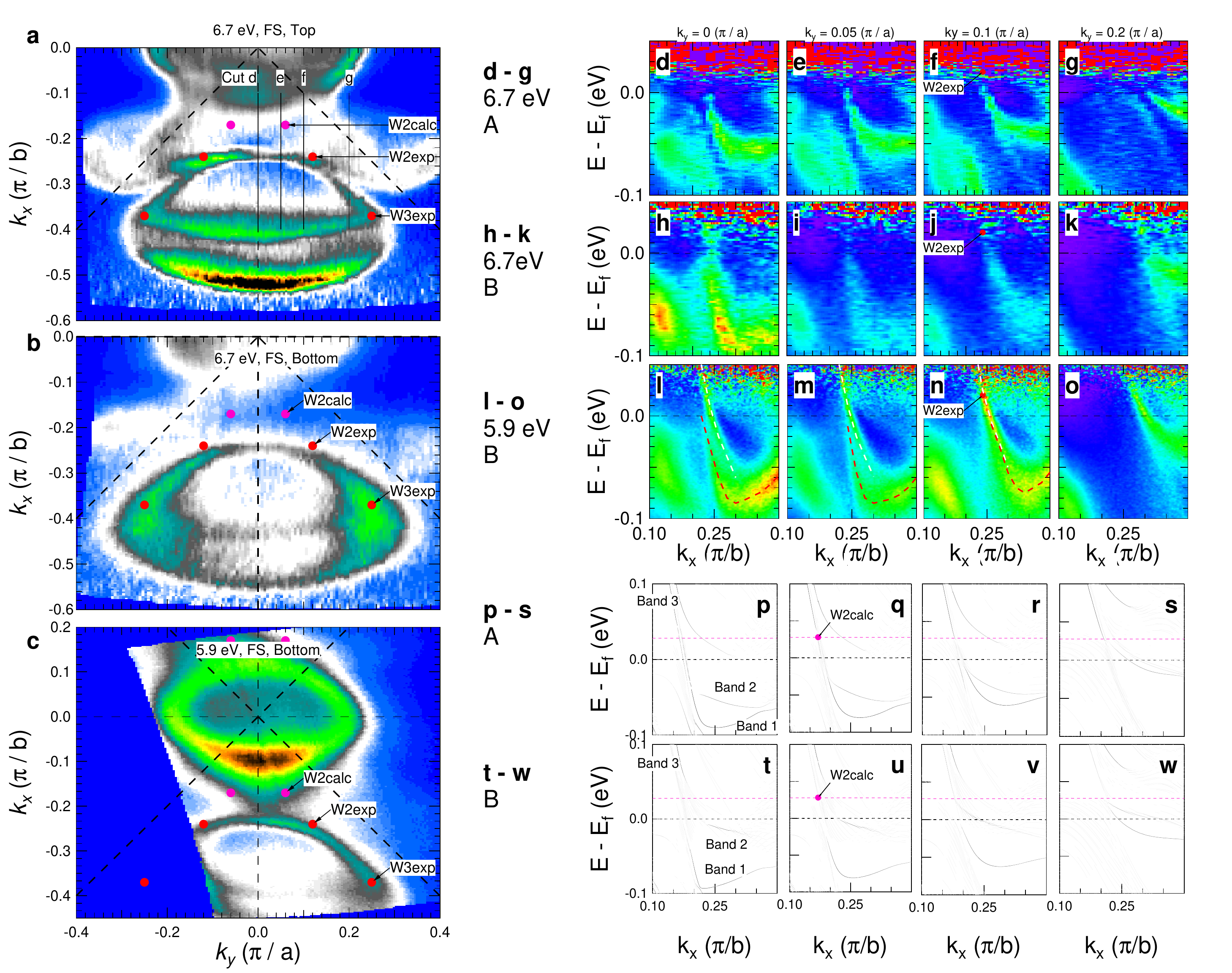}
	\caption{Identification of Weyl points and Fermi arcs from experimental data.
	{\bf a} Constant energy contour at E$_F$, measured by 6.7 eV photons for surface termination A. DFT predicted locations for Weyl points $W_2$ and measured Weyl points $W_2$, $W_3$ are marked as red and pink dots respectively. 
	{\bf b} The same panel as {\bf a} except for surface termination B.
	{\bf c} The same panel as {\bf b} except for using 5.9 eV photons.
	{\bf d - g} Energy dispersion for surface termination A along $k_y$ = 0, 0.05, 0.10 and 0.20 $\pi / a$. The projections of Weyl points $W_2$ are marked as dots. 
	{\bf h - k} The same panels as ({\bf d - g}) except for surface termination B.
	{\bf l - o} The same panels as ({\bf h - k}) except for using 5.9 eV photons. 
	{\bf p - s} Calculated band dispersion for surface termination A along cuts at $k_y$ = 0, 0.05, 0.10 and 0.20 $\pi / a$. Positions of $W_2$ are marked similarly as above.
	{\bf t - w} The same as ({\bf p - s}) except for surface termination B. Bands plotted with darker lines have more surface weights.}
	\label{fig3}
\end{figure*}

\begin{figure*}
	\centering
	\includegraphics[width=\columnwidth]{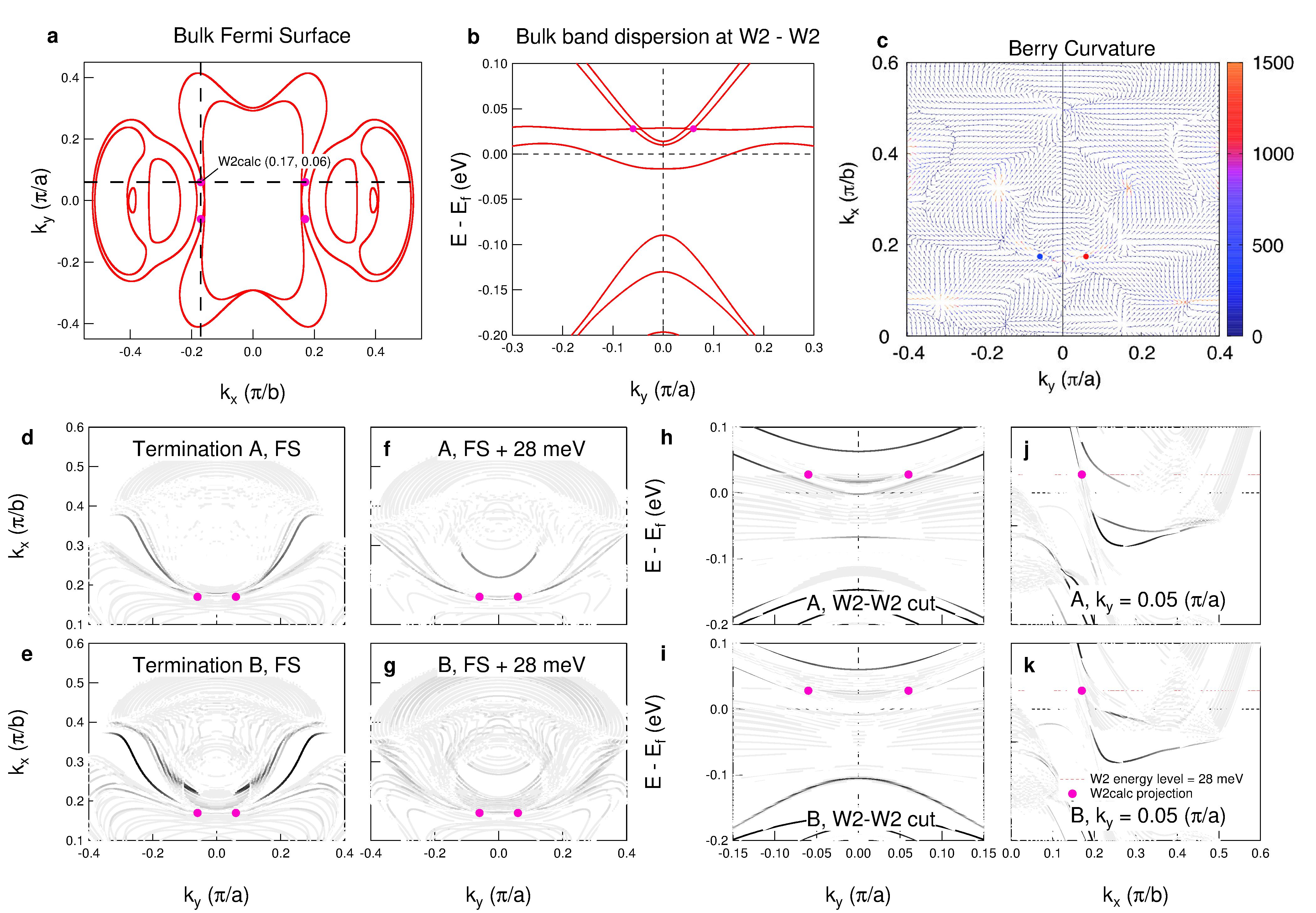}
	\caption{Results of DFT calculations.
	{\bf a} Calculated bulk Fermi surface of MoTe$_2$ for $k_z=0.6 \pi/c$ and projections of $W_2$ ($k_x$, $k_y$) = ($\pm$0.17 $\pi / b$, $\pm$0.06$\pi / a$) are marked with pink dots. 
	{\bf b} Bulk band dispersion along $W_2$-$W_2$ direction (the vertical dashed line in a). DFT predicted positions of $W_2$ ($k_y$, $E$) = ($\pm$0.06$\pi / a$, 0.028 eV) are marked. 
	{\bf c} The dominant contribution for the divergence of the Berry curvature $(\Omega^{DD}_{n,yz}, \Omega^{DD}_{n,zx})$ for the $n=N+1$ th band where $N$ is the number of electrons in the unit cell with $k_z=0$. Red and blue indicate different chiralities of the two Weyl points.
	{\bf d - g} Calculated constant energy contours of MoTe$_2$. Darker bands are surface bands and lighter bands are bulk bands. {\bf d, e} are at Fermi level for surface termination A and B. {\bf f, g} are at Fermi level + 28 meV of surface termination A and B, respectively. 
	{\bf h, i} Surface band dispersions of termination A and B along $W_2$-$W_2$ direction.
	{\bf j, k} Surface band dispersions of termination A and B along $k_y$ = 0.05 $\pi / a$ direction, which is very close to the $k_y$ position of $W_2$ (0.06 $\pi / a$). Positions of calculated Weyl points $W_2$ are marked and darker bands have more surface weights in {\bf d - k}.}
	\label{fig4}
\end{figure*}

Fig. \ref{fig4} is the DFT calculation of the band structure of MoTe$_2$. Fig \ref{fig4}a is the bulk Fermi surface for $k_z=0.6 \pi/c$ and calculated positions of four Weyl points are marked. The shapes of outermost electron and hole bands are very similar to our experiment result in Fig. \ref{fig2}b. Pink dots are projections of the calculated Weyl points on the $k_z=0$ plane from energy +28 meV above Fermi level, thus the electron band is not touching the two Weyl point projections. The surface weighted constant energy contours are shown in \ref{fig4}d - \ref{fig4}g. Fig. \ref{fig4}d and \ref{fig4}e are at Fermi surfaces of termination A and B, while \ref{fig4}f and \ref{fig4}g are at Fermi level + 28 meV, the DFT predicted energy of $W_2$. In the calculations, $W_2$ is not directly connected to another $W_2$ by surface states on the Fermi surface of termination A calculation while they are connected by weak and short surface states in termination B calculation. However, the $W_2$ points are connected by bulk electron bands in termination A. This is consistent with our experimental results shown in Fig. \ref{fig2}a-e. Fig. \ref{fig4}b is the bulk band dispersion at $W_2$-$W_2$ direction, as the vertical dashed line shown in \ref{fig4}a. The two $W_2$ points from DFT are right at the touching points of one hole band and one electron band. Fig. \ref{fig4}h and \ref{fig4}i show termination A and B surface band dispersions along the same direction as in Fig. \ref{fig4}b. The surface bands are to connect bulk states near the positions of the Weyl points. Fig \ref{fig4}j and \ref{fig4}k are termination A and B surface band dispersions along $k_y$ = 0.05 $\pi/a$ direction, as the horizontal dashed line shown in \ref{fig4}a. We also calculated the Berry curvature on Fermi surface. The bright points in Fig. \ref{fig4}c are possible singular points of the Berry curvature and DFT calculated $W_2$ points are marked in red and blue, indicating different chiralities of the Weyl points. The summary of energy and momentum locations of Weyl points based on calculations and experiment are provided in Table I.

\begin{table}
 \begin{center}
 \begin{tabular}{||c| c| c| c||} 
 \hline
  & $k_x\ (\pi/b)$ & $k_y\ (\pi/a)$ & $E\ $(meV) \\ [0.5ex] 
 \hline\hline
 $W_2$ DFT & $\pm 0.17$ & $\pm 0.06$ & 28 \\
 \hline
 $W_2$ Exp & $\pm 0.24$ & $\pm 0.12$ & 20 \\ 
 \hline
 $W_3$ Exp & $\pm 0.37$ & $\pm 0.25$ & 30 \\
 \hline
\end{tabular}
\end{center}
\label{weylLocationTable}
\caption{The locations $(k_x,k_y,E)$ of the Weyl points from DFT and ARPES. }
\end{table}

\section*{Methods}

\textbf{Sample growth} MoTe2 single crystals were grown out of a Te-rich binary melt using a Canfield crucible set(CCS) \cite{doi:10.1080/14786435.2015.1122248}. Mo and Te shots in a ratio of 1:9 were loaded into a 5ml CCS and sealed in a quartz tube under vacuum. The quartz ampoule was heated up to 1000C and kept at this temperature for a week. MoTe2 single crystals were isolated from Te flux by centrifuging. Different from most flux growths in which crystals precipitate while cooling from the homogenizing temperature, our growth was performed at a fixed temperature. Single crystals grown in this strategy have a RRR ~500 and MR~40,000\% at 2 K in an applied magnetic field of 100 kOe.

\textbf{Measurements} ARPES measurements were carried out using a laboratory-based system consisting of a Scienta R8000 electron analyzer and a a tunable VUV laser light source \cite{Jiang14RSI}. The data were acquired using a tunable VUV laser ARPES system, consisting of a Scienta R8000 electron analyzer, picosecond Ti:Sapphire oscillator and fourth harmonic generator. Angular resolution was set at $\sim$ 0.05$^\circ$ and 0.5$^\circ$ (0.005~\AA$^{-1}$ and 0.05 ~\AA$^{-1}$ ) along and perpendicular to the direction of the analyzer slit (and thus cut in the momentum space), respectively; and energy resolution was set at 1 meV. The size of the photon beam on the sample was $\sim$30 $\mu$m. Samples were cleaved \textit{in situ} at a base pressure lower than $1 \times 10^{-10}$ Torr. Samples were cooled using a closed cycle He-refrigerator and the sample temperature was measured using a silicon-diode sensor mounted on the sample holder. The energy corresponding to the chemical potential was determined from the Fermi edge of a polycrystalline Au reference in electrical contact with the sample.

\textbf{DFT calculations}
We first performed first-principles band structure calculations for bulk using the Perdew-Burke-Ernzerhof parametrization of the generalized gradient approximation~\cite{GGA} and the full-potential (linearized) augmented plane-wave plus local orbitals (FP-(L)APW+lo) method including the spin-orbit coupling as implemented in the \textsc{wien2k} code~\cite{Wien2k}. We employed the crystal structure determined by our experiment.
The muffin-tin radii for Mo and Te atoms, $r_{\mathrm{Mo}}$ and $r_{\mathrm{Te}}$, were set to 2.50 and 2.33 a.u., respectively. The maximum modulus for the reciprocal lattice vectors $K_{\mathrm{max}}$ was chosen so that $r_{\mathrm{Te}}K_{\mathrm{max}}$ = 8.00.
Next we constructed a tight-binding model consisting of Mo 4$d$ and Te 5$p$ orbitals, the parameters in which were extracted from the calculated band structure using the Wannier functions~\cite{Wannier1,Wannier2,Wannier3} without the maximal localization procedure.
Then we made the slab tight-binding model of finite layers, and obtained the band structures and Fermi surfaces. The Mo and Te states on the top or bottom two layers, i.e. the unit cells of surface A or B, are emphasized in each figure.
To identify the position of the Weyl points, we also calculated the Berry curvature for bulk structure using the tight-binding model obtained above. The dominant contribution for the divergence of the Berry curvature $(\Omega^{DD}_{n,yz}, \Omega^{DD}_{n,zx})$ presented in Eq. (30) of Ref.~\cite{Berrycalc} was calculated and shown in the figure.

\section*{Acknowledgements}
This work was supported by Center for Emergent Materials, an NSF MRSEC, under grant DMR-1420451 (theory and data analysis).  T. M. M. acknowledges funding from NSF-DMR-1309461 and would like to thank the 2015 Princeton Summer School for Condensed Matter Physics for their hospitality. H.C. received support from the Scientific User Facilities Division, Office of Basic Energy Sciences, US Department of Energy. ARPES data was acquired using Ames Laboratory spectrometer supported by the U.S. Department of Energy, Office of Science, Basic Energy Sciences, Materials Science and Engineering Division. Ames Laboratory is operated for the U.S. Department of Energy by Iowa State University under contract No. DE-AC02-07CH11358.

\section*{Author contributions}
N. T. and T. M. M. provided theoretical modeling and interpretation. J. Y. and Z. Z. grew the samples. M. O., M. S. and R. A. performed DFT and Berry phase calculations. H. C. performed crystal structure determination. L. H., Y. W. and D. M. performed ARPES measurements and support. L. H. analyzed ARPES data. The manuscript was drafted by L.~H., T.~M.~M., N.~T.  and A.~K. All authors discussed and commented on the manuscript.

\bibliography{MoTe2}

\end{document}